\documentstyle[epsbox]{article}
\begin{document}
\title{Interaction Effects on the Conductance in
One-Dimensional Systems \\-- Short-Range
Interaction\thanks{submitted to J. Phys. Soc. Jpn.}--\\}
\author{Arisato {\sc Kawabata}\thanks{E-mail:
arisato.kawabata@gakushuin.ac.jp} \\Department of Physics, Gakushuin
University, \\1-5-1 Mejiro, Toshima-ku, Tokyo 171}
\maketitle
\begin{abstract}
We investigate the effect of electron-electron
interactions on the conductance of quasi
one-dimensional systems without potential scattering. For a
finite temperature or system length, the short-range interaction
is not renormalized to 0, and it gives rise to a finite
correction to the conductance if we calculate it using   Kubo
formula. We show that this correction can be absorbed into the
renormalization of the chemical potential and that the properly
defined conductance to be observed in the experiments is equal to that of
non-interacting electrons.\\
\\
keyword: Tomonaga-Luttinger liquid, one-dimensional system
quantum wire, Coulomb interaction

\end{abstract}

\def\del{\delta}
\def\s{\sigma}
\def\S{{\mit \Sigma}}
\def\om{\omega}
\def\on{\omega_n}
\def\en{\varepsilon_n}
\def\em{\varepsilon_m}
\def\e{{\rm e}}
\def\d{{\rm d}}
\def\i{{\rm i}}
\def\kB{k_{\rm B}}
\def\kF{k_{\rm F}}
\def\vF{v_{\rm F}}
\def\eF{\varepsilon_{\rm F}}
\def\Del{{\mit \Delta}}
\def\Fi{{\mit \Phi}}
\def\G{{\cal G}}
\def\ua{\uparrow}
\def\da{\downarrow}
\def\&{&\!\!\!\!\!}

\section{Introduction}

Electron transport in one-dimensional interacting electron
system is not yet fully understood in spite of its simplicity.
Even in the absence of the potential scattering, until recently
it has been believed that the conductance is renormalized by
the the long-range interaction like $2Ke^2/h$, where $K<1$ for
repulsive interaction.\cite{Soly} The experiment by Tarucha et
al. ,\cite{Taru} however, indicates that the renormalization is
absent, and its has been shown by the present author that the
renormalization is the result of the incorrect definition of the
conductance.\cite{Kawa} 

He showed that the renormalization
can be derived with the use of the self-consistent field method:
When a electric potential is applied on the system, the
electrons are driven by the self-consistent potential which is
reduced because of the screening, while the current is defined
as the ratio of the current to the difference of the externally
applied potential
 between the ends of the system. This wrong
definition of the conductance leads to the apparent reduction
of the conductance. In fact, as has been pointed out by
Izuyama,\cite{Izuy} the conductance should be defined as the
ratio of the current to the difference of the self-consistent
potential between the ends of the system, because the
self-consistent potential which contains the contributions of
the polarization charge is the electric potential observed in
the experiments. Within the self-consistent field theory, which
is equivalent to the exact theory in the present case, the
response to the self-consistent potential is that of
non-interacting electron, and hence the correctly defined conductance
is not renormalized.

Here it should be noted that the above arguments are correct
only when the long-range parts of the interaction potential are
taken into account. It has been argued that the short-range
parts are renormalized to 0,\cite{Soly} but it is the case only
for 0K and infinitely long systems. In fact, in order to determine whether the
renormalization exists, the experiments have to be done at moderate
temperatures to avoid the effects of the scattering by
impurities or boundary irregularity. 

Thus it is important to investigate the effects of the
short-range part of the interaction on the conductance, and it is
the purpose of this paper.

\section{Correction to the Conductance}

We consider a one-dimensional interacting electron system described by 
the Hamiltonian
\begin{equation}
{\cal H} =\sum_{k, \s} \frac{\hbar^2k^2}{2m}a_{k, \s}^{\dagger}a_{k, \s}
+\frac{1}{L}\sum_{p, k, q}V(q) a_{p+q, \ua}^{\dagger}
a_{k-q, \da}^{\dagger}a_{k, \da}a_{p, \ua}\,.  \label{Hami}
\end{equation}

Suppose we apply an electric field $E(x)\e^{\del t}\,(\del\to +0)$ along the
system, then the current is given by\cite{Kubo} 
\begin{eqnarray}
I(x)&=&\int \sigma(x,x')E(x')\,\d x'\,,           \label{I(x)}\\
\sigma(x,x')&=&\lim_{\del\to +0}\int_0^\beta \d\lambda\int_0^\infty\d
t\,\langle j(x',-\i\hbar\lambda) j(x,t)\rangle\e^{\del t},   \label{sigma}
\end{eqnarray}

where $j(x,t)$ is the Heisenberg representation of the current operator.
It is easy to see that eq. (\ref{sigma}) can be rewritten in the form
\begin{eqnarray}
\sigma(x,x')=\lim_{\del\to +0}\frac{1}{\del}
\{G(x,x',\del)-G(x,x',0)\}\,, \hspace{0.5cm}       \label{sigma2} \\
G(x,x',\on)=-\int_0^{\hbar\beta} \langle \bar{j}(x,\tau)j(x')\rangle
\e^{\i\on\tau/\hbar}\,\d\tau\,,\hspace{0.5cm} \label{G}
\end{eqnarray}
with
\begin{equation}
\bar{j}(x,\tau)=\e^{{\cal H}\tau/\hbar}j(x) \e^{-{\cal H}\tau/\hbar} , 
\label{jth}
\end{equation}
where $\on=2\pi\kB Tn$, $n$ being an integer, and $G(x,x',\del)$ is
obtained by analytic continuation.

We let $G(q,\on)$ be the Fourier transform of $G(x,x',\on)$;
\begin{equation}
G(x,x',\on)=\int\frac{\d q}{2\pi}G(q,\on)\e^{\i q(x-x')}.        \label{Gq}
\end{equation}

\begin{figure}[hbt]
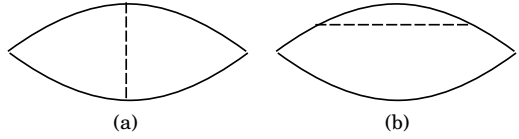

\begin{center}
\psbox[width=7cm]{FigAK.epsf}
\end{center}
\caption{Vertex correction (a), and self-energy correction (b) to
$G(q,\on)$. The solid and the dotted lines indicate the electron Green's
functions and the interactions, respectively.}
\label{cor}
\end{figure}

 In the lowest order in the short-range interaction, there are two
kinds of corrections to $G(q,\on)$,
namely, vertex correction (Fig.\ref{cor}(a)), and self-energy correction (Fig.
\ref{cor}(b)).

The vertex correction is given by

\begin{eqnarray}
\Del G_v(q,\on)=-2\hbar(\kB T)^2\sum_{\en,\em}\int\frac{\d k}{2\pi}
\int\frac{\d p}{2\pi} V(p-k)\left(\frac{e\hbar}{2m}\right)^2\hspace{1cm}
\nonumber \\
\times(2k+q)(2p+q)\G(k,\en)
G(k+q,\en+\on)\G(p,\em)\G(p+q,\em+\on)\,,      
\label{DG}
\end{eqnarray}

where $\G(k,\en)$ is the one-electron Green's function;
\begin{equation}
\G(k,\en)=\frac{1}{\i\en-\xi_k}\,,        \label{Gf}
\end{equation}
with $\en = (2n+1)\kB T$, $\xi_k=\hbar^2(k^2-\kF^2)/2m$ and $\kF$ being
the Fermi wave number. Using the standard technique of the thermal Green's
function method,\cite{Abri} we find that
\begin{eqnarray}
\Del G_v(q,\on)=-2\hbar\left(\frac{e\hbar}{2m}\right)^2\int
\frac{\d k}{2\pi}
\int\frac{\d p}{2\pi} V(p-k)\hspace{0.5cm}
\nonumber \\
\times(2k+q)(2p+q)\frac{f(\xi_{k+q})-f(\xi_k)}{\xi_{k+q}-\xi_k-\i\on}
  \,\frac{f(\xi_{p+q})-f(\xi_p)}{\xi_{p+q}-\xi_p-\i\on} \,, \nonumber \\
\label{DG2}
\end{eqnarray}
where $f(\xi)$ is the Fermi distribution function.

We assume that the spatial change of the electric field $E(x)$ is very
slowly, and we calculate $\Del G_v(q,\on)$ for $q \ll \kF$. Then, for $\kB T
\ll
\eF$, $\eF$ being the Fermi energy, in the right hand side of eq. (\ref{DG2})
only the wave number $k, p\approx\pm
\kF$ contribute to the integrals. Since we are interested in the correction
due to the short range part of $V(q)$, we consider only the contributions
from
$|k-p|\approx 2\kF$. Then we easily find that
\begin{equation}
\Del G_v(q,\on)=\left(\frac{e\vF}{\pi}\right)^2V(2\kF)
\frac{\hbar q^2}{(\hbar \vF q)^2+\on^2}       
\label{DG3}\,,
\end{equation}
and from eqs. (\ref{sigma2}) and (\ref{Gq}) that the vertex correction to
the conductivity is given by
\begin{eqnarray}
\Del \s_v(x,x')&=&-\lim_{\del\to +0}
\frac{e^2}{\pi^2\hbar}V(2\kF)
\int\frac{\del\,\e^{\i q(x-x')}}{(\hbar\vF q)^2+\del^2}
\frac{\d q}{2\pi} \nonumber \\
 &=&-\frac{2e^2}{h}\frac{V(2\kF)}{h \vF}\,.         
\label{Dsig}
\end{eqnarray}

Next we calculate the self-energy correction. The correction of
Fig.\ref{cor}(b) is given by
\begin{eqnarray}
\Del G_{s1}(q,\on)=2\hbar\kB
T\sum_{\en}\int\left(\frac{e\hbar}{2m}\right)^2 (2k+q)^2  \nonumber \\
\times\G^2(k,\en)\S(k)\G(k+q,\en+\on)\,\frac{\d k}{2\pi}\,,      
\label{Gs1}
\end{eqnarray}
 where $\S(k)$ is the self-energy part, which is independent of the energy
:
\begin{eqnarray}
\S(k)&=-&\kB T\sum_{\em}\int V(k-p)\G(p,\em)\frac{\d p}{2\pi}
\nonumber
\\ &=&-\int V(k-p)f(\xi_p)\frac{\d p}{2\pi}    \label{S(k)}\,.
\end{eqnarray}

Together with the contribution of the Feynman graph with the
self-energy correction to the other one-electron Green's function in 
Fig.\ref{cor}(b), we find that the self-energy correction to $G(q,\on)$ is
given by

\begin{eqnarray}
\Del G_s(q,\on)=-2\hbar\left(\frac{e\hbar}{2m}\right)^2\int (2k+q)^2
\hspace{4cm}\nonumber \\
\times\left[\frac{\{f(\xi_{k+q})-f(\xi_k)\}\{\S(k+q)-\S(k)\}}{(\xi_{k+q}-\xi_k-\i\on)^2}
+\frac{f'(\xi_{k+q})-f'(\xi_k)}{\xi_{k+q}-\xi_k-\i\on}\right]\frac{\d
k}{2\pi}\,, 
\label{DGs}
\end{eqnarray}
where the primes on $f(\xi)$ denote the derivative with respect to $\xi$.
 
The self-energy correction to the conductivity is obtained from eqs.
(\ref{sigma2}) and (\ref{Gq}) with $G(q,\on)$ replaced by $\Del G_s(q,\on)$
in eq. (\ref{DGs}). After some manipulations we find that
\begin{eqnarray}
\Del \s_s(x,x')
=\int\frac{e^2\hbar |k|}{m}[f'(\xi_k)\S'(k)-f''(\xi_k)\S(k)]
\frac{\d k}{2\pi}\,, \hspace{-0.5cm} \nonumber \\
       \label{Ds2}
\end{eqnarray}
where the primes indicate the derivative with respect to $\xi_k$.
It is easy to see that this equation can be written in the form
\begin{equation}
\Del \s_s(x,x')=\frac{4e^2}{\hbar}\int_0^\infty f'(\xi_k)\frac{\d \S(\xi_k)}
{\d k}\frac{\d k}{2\pi}\,.       
\label{Ds3}
\end{equation}

The derivative of the self-energy part is obtained from eq. (\ref{S(k)}):
\begin{equation}
\frac{\d \S(k)}{\d k}=-\int V(k-p)\frac{\d f(\xi_p)}{\d p}
\frac{\d p}{2\pi}\,,        \label{dS(k)}
\end{equation}
and for $\kB T\ll \eF$, it follows that
\begin{equation}
\Del \s_s(x,x')=\frac{2e^2}{h}\frac{2V(2\kF)}{h\vF}\,,        \label{Ds4}
\end{equation}
where we have neglected the contribution of the term with $V(q)$ for 
$q\ll 2\kF$.

Thus together with eq. (\ref{Dsig}) the correction to the conductivity to
the first order in the short-range part of the interaction is given by
\begin{equation}
\Del \s(x,x')=\frac{2e^2}{h}\frac{V(2\kF)}{h\vF}.        \label{Dst}
\end{equation}
Since the right hand side is independent of $x$ and $x'$, we find that the
correction to the conductance is also given by the right hand side of the
above equation.
  
\section{Renormalization of the Chemical Potential}

In this section we show that correction to the conductance obtained
in the preceding section can be interpreted as the one due to the
correction to the chemical potential difference at the ends of the sample.

The electric potential applied  on the system causes the deviation $\Del
n(x)$ of the electron density from its unperturbed value, and it gives rise to
the change in the local chemical potential.

We calculate $\Del n(x)$ as the response to the electric potential
$\Fi(x)\e^{\del t/\hbar} (\del
\to +0)$, with $E(x)=-\d\Fi(x)/\d x$, and we obtain
\begin{equation}
\Del n(x)=\lim_{\del\to 0}\int\Fi(q)R_0(q,\del)\e^{\i qx}\frac{\d
q}{2\pi}\,,
\label{Dn}
\end{equation}
where $\Fi(q)$ is the Fourier component of $\Fi(x)$ and $R_0(q,\del)$ is
the density response function:\cite{Kawa}
\begin{equation}
R_0(q,\del)=\frac{2e\vF q^2}{\pi[(\hbar\vF q)^2+\del^2]}\,. \label{R}
\end{equation}
Then we obtain
\begin{equation}
\Del n(x)=\frac{2e}{\pi\hbar\vF}\Fi(x)\,.  \label{Dn2}
\end{equation}
Note that the density of states per spin is $1/\pi\hbar\vF$. Hence the local
Fermi energy is given by
\begin{equation}
\Del \eF(x)=e\Fi(x)\,.        \label{DeF}
\end{equation}
Here the Fermi energy is measured from the bottom of the band.

On the other hand, the interaction correction to the chemical potential is
given by
\begin{equation}
\Del\mu = \Del\S(\kF)=-\int V(\kF-p)\Del f(\xi_p)\frac{\d p}{2\pi}\,,       
\label{Dmu}
\end{equation}
because the change of the self-energy have to be compensated by that of
the chemical potential in order to keep the electron density unchanged.  Here
we assume that the local relation between the Ferm energy and the chemical
potential holds, i.e.,
\begin{equation}
\Del\mu(x)=-\int V(\kF-p)\frac{\partial f(\xi_p)}{\partial\mu}
\Del\eF(x)\frac{\d p}{2\pi}\,, \label{Dmux}
\end{equation}
and then, neglecting the contributions of $V(q)$ for $|q|\ll 2\kF$, it follows
from eq. (\ref{DeF}) that 
\begin{equation}
\Del\mu(x)=-e\Fi(x)\frac{V(2\kF)}{2\pi\hbar\vF}\,.
\label{Dmux2}
\end{equation}

The increase of the current due to this correction to the chemical potential
difference is given by
\begin{eqnarray}
\Del I&=&\frac{2e^2}{h}\frac{\{\Del\mu(-\infty)-\Del\mu(\infty)\}}{-e}
\nonumber\\
 &=&\frac{2e^2}{h}\frac{V(2\kF)}{h\vF}\{\Fi(-\infty)-\Fi(\infty)\}\,.
\label{DI}
\end{eqnarray}
Comparing it with eq. (\ref{Dst}), we find that the correction to the
conductance due to the short-range parts of the interaction potential can be
explained in terms of the corrections to the chemical potential at the ends
of the system. Therefore, the correction to the conductance can be absorbed
into the renormalization of the chemical potential, i.e., the conductance is
not renormalized if we define it as the ratio of the current to the
difference of the renormalized chemical potential at the ends of the system.

\section{Discussion}

In the previous paper\cite{Kawa}, the present author has pointed out that
the renormalization  of the conductance due to the long-range part of the
interaction can be explained in terms of the renormalization of the
electric potential. Since the renormalized potential is the one observed
in the experiments, the
conductance should be  defined as the ratio of the current to the
difference of the renormalized potential at the ends of the sample. The
conductance defined in this way is not renormalized. 

In this paper, using Kubo formula, we showed that the short-range part of
the interaction gives a finite correction to the conductance and that it can
be ascribed to the renormalization of the chemical potential. As in the above,
the renormalized chemical potential is the one observed in the experiments,
and the conductance should be defined in terms of the renormalized
chemical potential. Then the conductance is not renormalized also in
this case. Although the present arguments are restricted to the lowest
order in the interaction potential, we can expect that it is the case to all
order.

Recently, there are some calculations on the effects of the Umklapp
scattering. In the Umklapp scattering, the total momentum of electrons is
not conserved, and we can expect a finite correction to the conductance.
From the point of view of the present theory, however, some contributions
to the correction obtained using Kubo formula  might be absorbed into the
renormalization of the chemical potential. Such contributions are not to be
observed in the experiments. 

 \section*{Acknowledgments}

The author is grateful to Prof. T. Ando, Prof. A. Shimizu, and Prof. K. Kitahara
for valuable discussions. This work 
is partly supported by Grant-in-Aid for Scientific Research 
on Priority Area "Mesoscopic Electronics: Physics and Technology"
from the Ministry of Education, Science, Sports and Culture.

\end{document}